\documentclass[lettersize,journal]{IEEEtran}
\usepackage{amsmath,amsfonts}
\usepackage{algorithmic}
\usepackage{algorithm}
\usepackage{array}
\usepackage[caption=false,font=normalsize,labelfont=sf,textfont=sf]{subfig}
\usepackage{textcomp}
\usepackage{stfloats}
\usepackage{url}
\usepackage{verbatim}
\usepackage{graphicx}
\usepackage{cite}
\usepackage{xcolor}
\usepackage{amsthm}

\newtheorem{theorem}{Theorem}
\newtheorem{lemma}{Lemma}

\newtheorem{corollary}{Corollary}

\newtheorem{remarkx}{Remark}
\newenvironment{remark}
  {\pushQED{\qed}\remarkx}
  {\popQED\endremarkx}

\newtheorem{assumption}{Assumption}

\begin{document}

\title{Event-triggered Control of Port-Hamiltonian Systems under Time-delay Communication}

\author{Ernesto Aranda-Escolastico$^{1}$, Leonardo J. Colombo$^{2}$, María Guinaldo$^{3}$, Antonio Visioli$^{4}$
        % <-this % stops a space
\thanks{$^{1}$ is with Department of Software and Systems Engineering, Universidad Nacional de Educación a Distancia (UNED), 28040 Madrid, Spain, {\tt\small earandae@issi.uned.es}}%

\thanks{$^{3}$ is with Centre for Automation and Robotics (CSICUPM), Ctra. M300 Campo Real, Km 0,200, Arganda del Rey - 28500 Madrid, Spain, {\tt\small leonardo.colombo@csic.es}}

\thanks{$^{3}$ is with the Computer Science and Automatic Control Department, UNED, Juan del Rosal 16, 28040, Madrid, Spain {\tt\small mguinaldo@dia.uned.es}}%

\thanks{$^{4}$ is with the Dipartimento di Ingegneria Meccanica e Industriale, University of Brescia, Brescia, Italy {\tt\small antonio.visioli@unibs.it}}%

}



\maketitle

\begin{abstract}
We study the problem of periodic event-triggered control of interconnected port-Hamiltonian systems subject to time-varying delays in their communication. In particular, we design a threshold parameter for the event-triggering condition, a sampling period, and a maximum allowable delay such that interconnected port-Hamiltonian control systems with periodic event-triggering mechanism under a time-delayed communication are able to achieve asymptotically stable behaviour. Simulation results are presented to validate the theory.
\end{abstract}


\section{Introduction}

\IEEEPARstart{T}{he} port-Hamiltonian framework appears as an alternative to Euler-Lagrange formalism to model nonlinear systems with dissipation \cite{maschke1993port,vanderSchaft1996,van2014port,sprangers2014reinforcement}, which makes it a natural candidate to describe many physical systems such as robots \cite{reyes2022virtual,farid2022finite}, quadrotors \cite{yuksel2014reshaping}, spacecrafts \cite{aoues2017modeling}, or microgrids \cite{schiffer2015stability,schiffer2016stability}. Essentially, port-Hamiltonian systems are based on a geometric structure that empathizes the importance of the total energy, the interconnection pattern, and the dissipation of the system. A key factor from the control point of view is this idea of interconnection since many control laws are implemented from an external device through external port variables \cite{ortega2002interconnection,cervera2007interconnection,ortega2008control,castanos2009asymptotic}. This achieves a special relevance with the development of networked control systems \cite{Hespanha2007} since, despite the obvious benefits of the communication networks, several challenges arise such as network delays, limited bandwidth, or loss of information \cite{Gupta2010a}. In this regard, several solutions have been proposed to deal with time-delays in port-Hamiltonian frameworks. 

In \cite{pasumarthy2009stability,kao2012stability}, time-delays appearing in a skew-symmetric form are studied. These results are improved in \cite{aoues2014robust} using the Wirtinger inequality. A larger class of delays that are not constrained to be skew-symmetric is studied in \cite{schiffer2015stability,schiffer2016stability}. In \cite{sun2016adaptive}, the delays are considered in the context of time-varying port-Hamiltonian systems. Recently, several works have focused on the effect of input delays when there exists an actuator saturation \cite{aoues2018sufficient,cai2021simultaneous,farid2022finite}. Despite these developments, the potential benefits of techniques specifically designed for networked control systems have not been studied for port-Hamiltonian systems. For this reason, in this paper, we present, to the best of our knowledge, the first event-triggered strategy for port-Hamiltonian frameworks. 

Event-triggered control \cite{Tabuada2007,Heemels2012,hu2015consensus,Aranda2020,yao2020event} is based on the idea of transmitting information only in the instants of time that the physical system demands it. In this way, communication resources can be saved and more efficiently used reducing congestion and delays. The traditional concept of event-triggered control involves an event-triggering condition that determines when an event is triggered and the information is sent through the network. Usually, this condition is continuously evaluated to obtain the exact triggering instant. However, this might yield several problems. On the one hand, it is necessary to guarantee a minimum inter-event time in the theoretical design to avoid Zeno behavior. On the other hand, the implementation of the continuous event-triggering condition might be problematic in digital platforms. For these reasons, periodic event-triggered frameworks \cite{Heemels2013,Yue2013,Aranda2016,hu2018resilient,luo2019periodic,Aranda2021} have been proposed with the aim to evaluate the event-triggering condition only in prefixed instants of time, combining advantages of event-triggered control and periodic control. This approach is adopted in this paper to port-Hamiltonian systems. Therefore, the main contribution of this paper is the design of the first periodic event-triggered control framework for port-Hamiltonian systems combined with the study of time-varying delays. This strategy emerges as a less demanding solution in terms of communication resources than previous approaches focused on time-delays.

The remainder of this paper is organized as follows. Preliminary concepts about port-Hamiltonian systems and the design of a periodic event-triggered port-Hamiltonian interconnection are introduced in Section 2. In Section 3, the main results about the stability of the new framework are obtained. In Section 4, the results are applied to a normalized pendulum and several simulations are obtained to show the benefits of the strategy. Finally, concluding remarks are provided in Section 5.

%Notation?

%\section*{Notation:}

\section{Generalities on Port-Hamiltonian Systems and Problem formulation}
\subsection{Port-Hamiltonian Systems}
We consider nonlinear port-Hamiltonian systems \cite{vanderSchaft1996}. The nonlinear control equations describing a port-Hamiltonian system $\Sigma$ are
    \begin{align}
        \dot{x}(t)&=\left(J(x(t))-R(x(t))\right)\left[\nabla H(x(t))\right]+G(x(t))u(t),\nonumber\\
        y(t)&=G(x(t))^\top\left[\nabla H(x(t))\right],\quad x(0)=x_0\label{eq:pHsystem}
    \end{align}
where $x\in\mathbb{R}^n$ is the state vector, $u\in\mathbb{R}^m$ is the control input, and $y\in\mathbb{R}^m$ is the system output. $J\in\mathbb{R}^{n\times n}$ is a skew-symmetric matrix, i.e. $J=-J^\top$, called the structure matrix and corresponds to the internal power-conserving structure of physical systems, such as oscillation between potential and kinetic energy, kinematic constraints, Kirchhoff's laws, transformers, etc. $R\in\mathbb{R}^{n\times n}$ is a positive matrix in presence of energy dissipation (due, for instance, to damping, viscosity, resistance, etc.), and $R=0$ in the lossless energy case. $G\in\mathbb{R}^{n\times m}$ is the input force matrices, so $G(x(t))u(t)$ denotes the generalized forces resulting from the control input $u$. Finally, $H$ denotes the energy balance
\begin{align*} 
    H(x(t))=&H(x(t_0))+\int_{t_0}^{t}y^\top(s)u(s)ds\\&-\int_{t_0}^{t}\left[\nabla H(x(s))\right]^\top R \left[\nabla H(x(s))\right] ds.
\end{align*}

The uncontrolled system, $\dot{x}=B(x)\nabla H(x)$ with $B(x):=J(x)-R(x)$ is assumed to have an isolated equilibrium point $x=x^{*}$. Since the change of coordinates $x\mapsto x-x^{*}$ will always move this equilibrium point to the origin, there is no loss of generality in taking $x^{*}=0$.

It is well known (see \cite{vanderSchaft1996} for instance) that $H$ is a storage function and $y^\top u$ is a supply rate with unit power. Hence, since $H$ is bounded from below, system \eqref{eq:pHsystem} is said to be passive as it is dissipative with respect to the supply rate $y^\top u$, that is, the amount of energy of the system at time $t$ is equal to the amount of energy at time $t_0$ increased (or decreased) by the energy supplied (or removed) by the port variables and decreased by the dissipated energy. %\textcolor{red}{Mirar paper Seuret Asymp. Stab.}.

It is well-known that port-Hamiltonian systems are composable, i.e., the interconnection of two port-Hamiltonian systems $\Sigma_1$ and $\Sigma_2$ through a power-conserving structure yields a dynamical system $\Sigma_{12}$ which is again a port-Hamiltonian system. Note that this is a useful property since port-Hamiltonian controllers can be designed to, for instance, achieve asymptotic stability in set-points which are not a minimum of the Hamiltonian \cite{van2014port}. In the next sections, we study this interconnection when it is made through a communication network, and consequently, limited to its capabilities. In particular, in this work, we study the use of periodic event-triggering mechanisms to reduce communications between port-Hamiltonian systems and the effect of time-delays due to the network communication.
\subsection{Problem description: Periodic event-triggered interconnection}

Consider two input-to-state port-Hamiltonian systems $\Sigma_1$ and $\Sigma_2$ such that \begin{align}
        \dot{x}_i(t)&=\left(J_i(x_i)-R_i(x_i))\right)\left[\nabla H_i(x_i(t))\right]+G_i(x_i) u_i(t) \nonumber\\
        y_i(t)&=G_i^\top(x_i)\left[\nabla H_i(x_i(t))\right],\,\hbox{ for }i=1,2.\label{eq:pHsystem_i}
    \end{align} %\textcolor{purple}{where $J,G$ and $R$ are constants matrices.}

Therefore, an interconnected system $\Sigma_{12}$ can be described by \begin{equation} \label{eq:pHsystem_12}
    \begin{aligned}
        \begin{bmatrix}\dot{x}_1(t)\\\dot{x}_2(t)\end{bmatrix}=&\begin{bmatrix}B_1(x_1(t)) & 0 \\ 0 & B_2(x_2(t))\end{bmatrix}\begin{bmatrix}\nabla H_1(x_1(t)) \\ \nabla H_2(x_2(t))\end{bmatrix}
        \\
        &+\begin{bmatrix}G_1(x_1(t)) & 0 \\ 0 & G_2(x_2(t))\end{bmatrix}\begin{bmatrix}u_1 (t)\\ u_2 (t)\end{bmatrix}
    \end{aligned},\end{equation}where we are denoting by $B_i(x_i):=J_i(x_i)-R_i(x_i)$, $i=1,2$.

\begin{figure}
    \centering
    \includegraphics[width=0.8\linewidth]{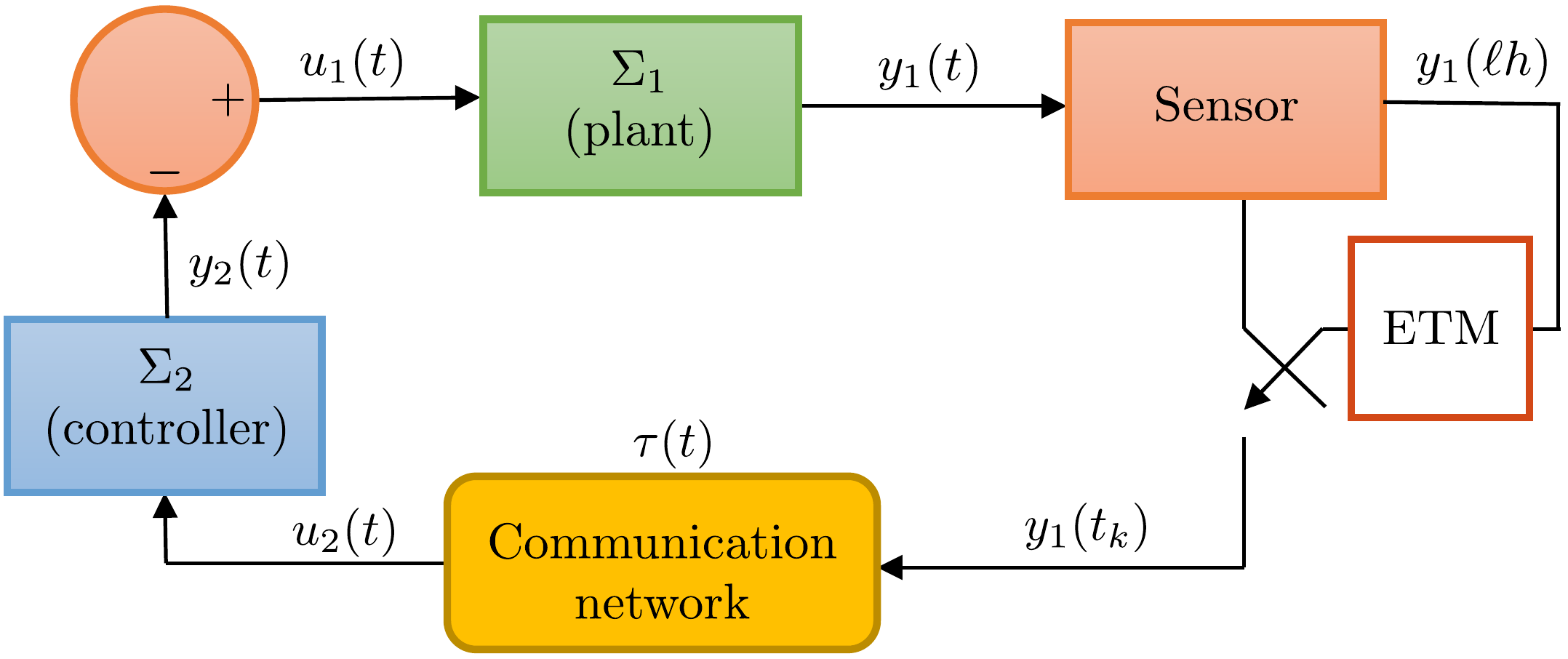}
    \caption{Block diagram of the event-triggered interconnected port-Hamiltonian system.}
    \label{fig:Block_diagram}
\end{figure}

Let us describe the feedback interconnection between $\Sigma_1$ and $\Sigma_2$ as drawn in Figure \ref{fig:Block_diagram}. In this scheme, we assume that the output $y_1$ is sampled with period $h>0$, i.e., the sampling sequence is described by the set $S_{1}=\{0,h,2h,...,\ell h\}$ for $\ell\in\mathbb{N}$. Whether or not the data should be transmitted over a communication network is determined by an event-triggering mechanism at the output of each port-Hamiltonian system. Then, the transmission sequence from $\Sigma_1$ to $\Sigma_2$ is described by $S_{2}=\{t_1,t_2,...,t_k\}\subseteq S_1$ for $k\in\mathbb{N}$. We  also consider that the communication network induces a time-varying delay $\tau(t)$ which satisfies $0< \tau_m \leq \tau(t) \leq \tau_M$, where $\tau_m$ and $\tau_M$ are the minimum and the maximum delay, respectively. Finally, the inputs $u_1$ and $u_2$ are generated by a zero-order-hold (ZOH) with the holding time $t\in[t_k+\tau(t_k),t_{k+1}+\tau(t_{k+1}))$. Consequently, denoting by $\mathcal{I}$ the identity matrix, we can write
\begin{equation} \label{eq:u}
\begin{bmatrix} u_1(t) \\ u_2(t) \end{bmatrix} = \begin{bmatrix} 0 & -\mathcal{I} \\ \mathcal{I} & 0 \end{bmatrix}\begin{bmatrix} y_1(t_k) \\ y_2(t) \end{bmatrix},
\end{equation}where for simplicity we denote $\hat{y}_1(t):=y_1(t_k)$ for $t\in[t_k+\tau(t_k),t_{k+1}+\tau(t_{k+1}))$.

\vspace{.2cm}

\begin{remarkx}
  To simplify the scheme and focus on the main problem, we consider a sampler and an event-triggering mechanism at the output of the  subsystem $\Sigma_1$. However, the procedure in \cite{Aranda2018} can be followed to obtain a dual-rate framework with two samplers, two event-triggering mechanisms and their respective delays.\hfill$\diamond$
\end{remarkx}

Now, let us express the transmission instants such as
\begin{equation} \label{eq:ETM}
    t_k=\inf \{\ell h: \ell\in\mathbb{N}, \ell h > t_{k-1}, \mathcal{C}(e(\ell h),y_1(\ell h))\geq 0\},
\end{equation}
where $e(t)$ is the error vector
\begin{equation} \label{eq:error}
    e(t)=\hat{y}_1(t)-y_1(t),
\end{equation}
for $t\in[t_k+\tau(t_k),t_{k+1}+\tau(t_{k+1}))$, and
\begin{equation}
    \mathcal{C}(e(t),y_1(t)) = e^\top(t)\Omega e(t) - \sigma y_1^\top (t) \Omega y_1(t),
\end{equation}being $\Omega\in\mathbb{R}^{m\times m}$ a positive definite weighting matrix to be designed and $\sigma\geq 0$ is a given parameter that regulates the number of events.

\vspace{.2cm}

\begin{remark}
Note that if $\sigma$ is small, then the event-triggering mechanism \eqref{eq:ETM} gets more sensitive to the output change and transmits the signal more frequently, which makes the controller \eqref{eq:u} more similar to a periodic controller. Note also that in the limit case, i.e., $\sigma\to0$, the system is transformed into a sampled-data system. On the contrary, if $\sigma$ is enlarged, less events are triggered, but properties such as stability might be compromised.\end{remark}

For a detailed timing analysis, we divide the holding interval $[t_k+\tau(t_k),t_{k+1}+\tau(t_{k+1}))$ into sampling intervals $[\ell h+\tau(\ell h),(\ell+1)h +\tau((\ell+1)h))$ and define a piece-wise function $\delta(t)=t-\ell h$ satisfying $0< \tau_m \leq \delta(t) \leq \tau_M +h = \delta_M$.

Since the event-triggering condition is evaluated in each $\ell h=t-\delta(t)$, it is convenient to write
\begin{equation} \label{eq:error2}
    y_1(t_k)=y_1(t-\delta(t)) + e(t-\delta(t)).
\end{equation}
Combining now \eqref{eq:pHsystem_12} and \eqref{eq:u} with  \eqref{eq:error2}, we obtain
\begin{equation} \label{eq:pHsystem_12_2}
    \begin{aligned}
        &\begin{bmatrix}\dot{x}_1(t)\\\dot{x}_2(t)\end{bmatrix}=\begin{bmatrix}B_1(x_1(t)) & 0 \\ 0 & B_2(x_2(t))\end{bmatrix}\begin{bmatrix}\nabla H_1(x_1(t)) \\ \nabla H_2(x_2(t))\end{bmatrix}
        \\
        &+\begin{bmatrix}0 & -G_1(x_1(t)) \\ G_2(x_2(t)) & 0\end{bmatrix}
        \\
        &\times\begin{bmatrix}y_1(t-\delta(t)) + e(t-\delta(t)))\\ y_2 (t)\end{bmatrix}
    \end{aligned}
\end{equation}
for $t\in[t_k+\tau(t_k),t_{k+1}+\tau(t_{k+1}))$. 

Next, if we replace the output using \eqref{eq:pHsystem_i} in \eqref{eq:pHsystem_12_2}, then
\begin{equation} \label{eq:pHsystem_12_final}
    \begin{aligned}
        &\begin{bmatrix}\dot{x}_1(t)\\\dot{x}_2(t)\end{bmatrix}=\begin{bmatrix}B_1(x_1(t)) & 0 \\ 0 & B_2(x_2(t))\end{bmatrix}\begin{bmatrix}\nabla H_1(x_1(t)) \\ \nabla H_2(x_2(t))\end{bmatrix}
        \\
        &+\begin{bmatrix}0 & -G_1(x_1(t))G_2^\top(x_2(t)) \\ 0 & 0\end{bmatrix}\begin{bmatrix}\left[\nabla H_1(x_1(t))\right]\\ \left[\nabla H_2(x_2(t))\right]\end{bmatrix}
        \\
        &+ \begin{bmatrix}0 & 0 \\ G_2(x_2(t))G_1^\top(x_1(t)) & 0\end{bmatrix}\begin{bmatrix}\left[\nabla H_1(x_1(t-\delta(t)))\right]\\ \left[\nabla H_2(x_2(t-\delta(t)))\right]\end{bmatrix}\\&+\begin{bmatrix}0 \\ G_2(x_2(t))\end{bmatrix}e(t-\delta(t)).
    \end{aligned}
\end{equation}
Denote the state of the system $\Sigma_{12}$ by $\xi=\begin{bmatrix}x_1 & x_2\end{bmatrix}^\top$ and its total energy by $\mathcal{H}(\xi)=H_1(x_1)+H_2(x_2)$. For the sake of simplicity, we refer in the following by $\mathcal{H}_{t}$ to indicate that $\mathcal{H}$ is taken at time $t$, i.e. $\mathcal{H}(\xi(t))=\mathcal{H}_t$. Thus, \eqref{eq:pHsystem_12_final} can be written as
\begin{equation} \label{eq:xi}
    \dot{\xi}(t)=A\nabla\mathcal{H}_t+A_d\nabla\mathcal{H}_{t-\delta(t)}+A_ee(t-\delta(t)),
\end{equation}
where $A=\begin{bmatrix}B_1(x_1(t)) & -G_1(x_1(t))G_2^\top(x_2(t))  \\ 0 & B_2(x_2(t))\end{bmatrix}$,\newline $A_d=\begin{bmatrix}0 & 0 \\ G_2(x_2(t))G_1^\top(x_1(t)) & 0\end{bmatrix}$ and $A_e=\begin{bmatrix}0 \\ G_2(x_2(t))\end{bmatrix}$.

%\textcolor{red}{Abuso notacion sin dependencias en A?} 

Note that \eqref{eq:xi} describes a system formed by two port-Hamiltionan subsystems with a periodic event-triggered interconnection and perturbed with time-delays. However, the port-Hamiltonian structure is no longer preserved due to the presence of sampled data and delays in $\nabla\mathcal{H}_{t-\delta(t)}$, and the stability cannot be concluded only from the properties of the Hamiltonian $\mathcal{H}$. Instead of that, we consider \eqref{eq:xi} as an \textit{interconnected time-delayed port-Hamiltonian system}, i.e. a nonlinear time-delayed system in a perturbed Hamiltonian form, and use Lyapunov-Krasovskii theory
(see \cite{fridman2014tutorial} for instance) to take into account the periodic event-triggered scheme and the delays. For the further analysis, two assumptions over the system \eqref{eq:xi} are also necessary:

\vspace{.2cm}

\begin{assumption} \label{assu1}
The time-delayed port-Hamiltonian system \eqref{eq:xi} posses an equilibrium point $\xi_e=0$.
\end{assumption}

\vspace{.2cm}

\begin{assumption} \label{assu2}
The Hamiltonian $\mathcal{H}$ is regular and positive definite around $\xi_e$ (i.e., $\mathcal{H}>0$, $\nabla\mathcal{H}(\xi_e)=0$ and  $\nabla\mathcal{H}(\xi)\neq 0$ for $\xi\neq\xi_e$ around $\xi_e$).
\end{assumption}

Note that Assumption \ref{assu2} implies that the port-Hamiltonian system is locally asymptotically stable in absence of event-triggered transmissions and time-delays with Lyapunov function $\mathcal{H}$.

Therefore, the following problem is established:

\textbf{Problem statement}: Given the interconnected time-delayed port-Hamiltonian system \eqref{eq:xi} under Assumptions \ref{assu1}-\ref{assu2}, design the sampling period $h$, the maximum allowable delay $\tau_M$ and the threshold parameter $\sigma$, such that \eqref{eq:xi} with periodic event-triggering mechanism \eqref{eq:ETM} is asymptotically stable.

\section{Control design}
In this section, the stability of the interconnected time-delayed port-Hamiltonian system \eqref{eq:xi} with event-triggered mechanism \eqref{eq:ETM} and under time-varying delays is studied. First, the local asymptotic stability is proved. Then, the conditions of global asymptotic stability are proposed and the particular case of linear port-Hamiltonian systems is studied. For this purpose, we first introduce the following technical results: the Lyapunov-Krasovskii Theorem and the Wirtinger Inequality. 

\vspace{.2cm}

Denote by $C[a,b]$ the space of continuous functions on $[a,b]$, $||\cdot||_{C}$ the norm on the space $C[a,b]$, and $\mathcal{B}(\Gamma)$ the space of bounded sets on $\Gamma\subset C[a,b]$. 

\vspace{.2cm}

\begin{lemma}[adapted from {\cite{fridman2014tutorial}}, Theorem 1] \label{Theo:LK}
Suppose $f:\mathbb{R}\times C[-\theta,0]\to\mathbb{R}^n$ maps $\mathbb{R}\times\mathcal{B}(C[-\theta,0])$ into $\mathcal{B}(\mathbb{R}^n)$ and that $\mu$, $\nu$ and $\rho:\mathbb{R}_+ \to\mathbb{R}_+$ are continuous nondecreasing functions, $\mu(s)$ and $\nu(s)$ are positive for $s>0$, and $\mu(0)=\nu(0)=0$. The trivial solution of $\dot{x}(t)=f(t,x(t))$ is uniformly stable if there exists a continuous functional $V:\mathbb{R}:\times C[-\theta,0]\to\mathbb{R}_+$, called Lyapunov-Krasovskii functional, which is positive-definite, i.e.
\begin{equation} \label{eq:L1}
    \mu(\|\phi(0)\|) \leq V(t,\phi) \leq \nu(\|\phi\|_C),
\end{equation}and such that its derivative along $\dot{x}(t)=f(t,x(t))$ is non-positive in the sense that
\begin{equation} \label{eq:L2}
    \dot{V}(t,\phi) \leq -\rho(\|\phi(0)\|).
\end{equation}
If $\rho(s)>0$ for $s>0$, then the trivial solution is uniformly asymptotically stable. If in addition $\lim_{s\to\infty}\mu(s)=\infty$, then it is globally uniformly asymptotically stable.
\end{lemma}

%\textcolor{red}{habria que definir propiamente que es un "Lyapunov-Krasovskii''functional}

%\textcolor{purple}{Habria que incluir un appedice con el S-procedure en lugar de referenciar a [7] o a continuacion del Lema 2, lo mismo para comparison lemma}

\vspace{.2cm}

\begin{lemma} [adapted from \cite{Seuret2013a}, Corollary 4] \label{lem2}
Consider a given matrix $Q\succ 0$. Then, for any continuous function $\omega:[a,b]\to\mathbb{R}^n$ the following inequality holds:
    \begin{align}
        \int_{a}^{b}\omega^{\top}(\beta)&Q\omega(\beta)d\beta\geq\frac{3}{b-a}\Phi^\top Q\Phi \label{eq:Wir}\\ &+\frac{1}{b-a} \left(\int_{a}^{b}\omega(\beta)d\beta\right)^{\top}Q\left(\int_{a}^{b}\omega(\beta)d\beta\right)\nonumber
    \end{align}
where $\Phi=\int_a^b \omega(s)ds - \frac{2}{b-a}\int_a^b\int_s^a\omega(r)drds$.
\end{lemma}

In order to fulfill the problem under study, the following theorem can be stated.

\vspace{.2cm}

\begin{theorem} \label{theo1}
    For given positive scalars $\sigma$, $h$ and $\tau_M$, the interconnected delayed port-Hamiltonian system \eqref{eq:xi} with event-triggered mechanism \eqref{eq:ETM} and under Assumptions \ref{assu1}-\ref{assu2} is asymptotically stable, if there exist matrices $P$, $Q$ and $\Omega$ - of appropriate dimensions - such that
    \begin{equation} \label{eq:P}
       (i)\,\, P \succ 0,\,\, (ii)\,\,  Q \succ 0,\,\ (iii)\,\,  \Omega \succ 0 \end{equation}
    \begin{equation} \label{eq:Theo}
        \Xi = \begin{bmatrix} \Xi_{11} & \star & \star & \star \\ \Xi_{21} & \Xi_{22} & \star & \star \\ 6Q & 6Q & -12Q & \star \\ \Xi_{41} & \Xi_{42} & 0 & \Xi_{44}\end{bmatrix}\prec0,
    \end{equation}is feasible around a neighborhood of $\xi_e=0$ and with
    \begin{align*}
        \Xi_{11} &= \frac{1}{2}\left(A+A^\top\right) + A^\top\nabla^2\mathcal{H}_t^\top P + P\nabla^2\mathcal{H}_t A\\ &+ \delta_M^2 A^\top \nabla^2\mathcal{H}_t^\top Q\nabla^2\mathcal{H}_t A - 4Q,\\
        \Xi_{21} &= A_d^\top\nabla^2\mathcal{H}_t^\top P + \frac{A_d^\top}{2} \\& + \delta_M^2 A_d^\top\nabla^2\mathcal{H}_t^\top Q \nabla^2\mathcal{H}_t A - 2Q,\\
        \Xi_{22} &= \delta_M^2 A_d^\top \nabla^2\mathcal{H}_t^\top Q \nabla^2\mathcal{H}_t A_d - 4Q + \sigma \mathcal{G}^\top \Omega \mathcal{G},\\
        \Xi_{41} &= A_e^\top\nabla^2\mathcal{H}_t^\top P + \frac{A_e^\top}{2} + \delta_M^2 A_e^\top \nabla^2\mathcal{H}_t^\top Q \nabla^2\mathcal{H}_t A,\\
        \Xi_{42} &= \delta_M^2 A_e^\top \nabla^2\mathcal{H}_t^\top Q \nabla^2\mathcal{H}_t A_d,\\
        \Xi_{44}&=\delta_M^2 A_e^\top \nabla^2\mathcal{H}_t^\top Q \nabla^2\mathcal{H}_t A_e - \Omega,\,    \mathcal{G}= \begin{bmatrix}
        G_1^\top & 0
        \end{bmatrix}.
    \end{align*}
Besides, if \eqref{eq:P}-\eqref{eq:Theo} is feasible for all $\xi\in\mathbb{R}^{n_1\times n_2}$, then the system is globally asymptotically stable.
\end{theorem}

\begin{proof}
Consider the functional
\begin{equation} \label{eq:LKF}
    V=V_1+V_2,
\end{equation}
where
\begin{align}
    V_1&=\mathcal{H}_t+\nabla\mathcal{H}^\top_t P \nabla\mathcal{H}_t,
    \\
    V_2&=\delta_M\int_{-\delta_M}^{0}\int_{t+s}^{t}\frac{d}{dr}\left(\nabla\mathcal{H}^\top_r\right) Q \frac{d}{dr}\left(\nabla\mathcal{H}_r\right)drds.
\end{align}
Since $\mathcal{H}$ is positive definite around the equilibrium point by Assumption \ref{assu2} and considering \eqref{eq:P} $(i)$ and $(ii)$, then \eqref{eq:LKF} is an admisible Lyapunov-Krasovskii Function satisfying \eqref{eq:L1}. The derivative of $V_1$ along the trajectories \eqref{eq:xi} is
\begin{equation} \label{eq:dV1}
    \begin{aligned}
        \dot{V}_1 =& \nabla\mathcal{H}^\top_t \dot{\xi}(t) + \nabla\mathcal{H}^\top_t P \nabla^2\mathcal{H}_t \dot{\xi}(t) 
        \\
        &+ \dot{\xi}^\top(t)\nabla^2\mathcal{H}_t P \nabla\mathcal{H}_t
        \\
        =& \nabla\mathcal{H}^\top_t\left(\frac{1}{2}(A+A^\top)+A^\top\nabla^2\mathcal{H}^\top_tP+P\nabla^2\mathcal{H}_tA\right)\nabla\mathcal{H}_t
        \\
        &+\nabla\mathcal{H}^\top_{t-\delta(t)}\left(\frac{A_d^\top}{2}+A_d^\top\nabla^2\mathcal{H}^\top_t P\right)\nabla\mathcal{H}_t
        \\
        &+e^\top(t-\delta(t))\left(\frac{A_e^\top}{2}+A_e^\top\nabla^2\mathcal{H}^\top_tP\right)\nabla\mathcal{H}_t
        \\
        &+\nabla\mathcal{H}^\top_t\left(\frac{A_d}{2}+P\nabla^2\mathcal{H}_t A_d\right)\nabla\mathcal{H}_{t-\delta(t)}
        \\
        &+\nabla\mathcal{H}^\top_t\left(\frac{A_e}{2}+P\nabla^2\mathcal{H}_t A_e\right)e(t-\delta(t)),
    \end{aligned}
\end{equation}
while the derivative of $V_2$ along \eqref{eq:xi} is
\begin{equation} \label{eq:dV2}
    \begin{aligned}
        \dot{V}_2 = &\delta_M^2 \dot{\xi}^\top\nabla^2\mathcal{H}^\top_t Q \nabla^2\mathcal{H}_t\dot{\xi} 
        \\
        &- \delta_M\int_{t-\delta_M}^{t}\frac{d}{ds}\nabla\mathcal{H}^\top_sQ\nabla\mathcal{H}_sds.
    \end{aligned}
\end{equation}
Applying now the Wirtinger inequality for time-delay systems (Lemma \ref{lem2}), the integral term \eqref{eq:dV2} is bounded as
\begin{equation} \label{eq:app_wirt}
    \begin{aligned}
            - \delta_M&\int_{t-\delta_M}^{t}\frac{d}{ds}\nabla\mathcal{H}^\top_sQ\nabla\mathcal{H}_sds \leq
            \\
            &\begin{bmatrix}\nabla\mathcal{H}_t-\nabla\mathcal{H}_{t-\delta(t)} \\ \nabla\mathcal{H}_t+\nabla\mathcal{H}_{t-\delta(t)} - \frac{2}{\delta(t)}\int_{t-\delta(t)}^t\nabla\mathcal{H}_sds \end{bmatrix}^\top \begin{bmatrix}Q & 0 \\ 0 & 3Q\end{bmatrix}\\ & \times \begin{bmatrix}\nabla\mathcal{H}_t-\nabla\mathcal{H}_{t-\delta(t)} \\ \nabla\mathcal{H}_t+\nabla\mathcal{H}_{t-\delta(t)} - \frac{2}{\delta(t)}\int_{t-\delta(t)}^t\nabla\mathcal{H}_sds \end{bmatrix}.
    \end{aligned}
\end{equation}
Now, equations \eqref{eq:dV1}, \eqref{eq:dV2} and \eqref{eq:app_wirt} are combined, and the S-procedure \cite{yakubovic1977s} is used to take into account the knowledge about the event-triggering mechanism \eqref{eq:ETM}, i.e., the quadratic form $-e^\top(t)\Omega e(t) + \sigma \nabla\mathcal{H}_{t-\delta(t)}^\top \mathcal{G}^\top \Omega \mathcal{G} \nabla\mathcal{H}_{t-\delta(t)} \leq 0$, with $\Omega$ satisfying \eqref{eq:P} $(iii)$, is added to the right of $\dot{V}$ such that
\begin{align*}
    \dot{V} &= \dot{V}_1 + \dot{V}_2\\& \leq \begin{bmatrix}\nabla\mathcal{H}_t \\ \nabla\mathcal{H}_{t-\delta(t)} \\  \frac{1}{\delta(t)}\int_{t-\delta(t)}^t\nabla\mathcal{H}_sds \\ e(t-\delta(t)\end{bmatrix}^\top \Xi \begin{bmatrix}\nabla\mathcal{H}_t \\ \nabla\mathcal{H}_{t-\delta(t)} \\  \frac{1}{\delta(t)}\int_{t-\delta(t)}^t\nabla\mathcal{H}_sds \\ e(t-\delta(t)\end{bmatrix}\\&=\psi^\top(t)\Xi\psi(t).
\end{align*}
Since $\Xi<0$ by \eqref{eq:Theo}, a constant $c>0$ exists such that $\dot{V}\leq -c \|\psi(t)\|\leq -c\|\nabla\mathcal{H}_t\|$. Since $H$ is assumed regular around $\xi_e$ by Assumption \ref{assu2}, $\|\nabla\mathcal{H}_t\|$ is a continuous positive definite function in a neighborhood of the origin. Consequently, by the comparison lemma (Lemma IV.1, \cite{angeli2000characterization}), a class $\mathcal{K}_\infty$ function $\kappa$ exists such that $\kappa(\|\xi(t)\|) \leq \|\nabla\mathcal{H}_t\|$ in a neighborhood of the $\xi_e$. Thus, \eqref{eq:L2} is satisfied, and the time-delayed port-Hamiltonian system \eqref{eq:xi} is asymptotically stable.
\end{proof}

\vspace{.1cm}

Note that \eqref{eq:Theo} is generally state-dependant due to the term $\nabla^2\mathcal{H}_t$ and the matrices $A$, $A_d$ and $A_e$. Consequently, the feasibility of \eqref{eq:Theo} might be difficult to prove.  To overcome this issue, a polytopic approach can be followed similarly as in  \cite{aoues2014robust,schiffer2016stability,aoues2018sufficient}. The objective is to transform \eqref{eq:P}-\eqref{eq:Theo} into a set of linear matrix inequalities (LMIs), which can be efficiently solved. 
Previously, two assumptions are necessary.

\vspace{.2cm}

\begin{assumption} \label{assu3}
    $J_i$, $R_i$ and $G_i$ are constant matrices for $i=1,2$.
\end{assumption}

\vspace{.2cm}

\begin{assumption} \label{assu4}
   The Hessian $\nabla^2\mathcal{H}_t$ is embedded into a polytope $\mathbb{P}$.
\end{assumption}

%Then, the following corollary can be stated.

\vspace{.2cm}

\begin{theorem} \label{theo2}
    Under Assumptions \ref{assu3}-\ref{assu4}, the asymptotic stability conditions \eqref{eq:P}-\eqref{eq:Theo} are satisfied if they are satisfied on the set of vertices of the polytope $[0,\delta_M]\times\mathbb{P}$.
\end{theorem}
\begin{proof}
First, let us show that \eqref{eq:Theo} is affine with respect to $\nabla^2\mathcal{H}_t$. Applying the Schur complement \cite{zhang2006schur} over \eqref{eq:Theo}, it is obtained that $\Xi\prec0$ if and only if $\Theta\prec0$ where
\begin{equation} \label{eq:Theo_Schur}
    \Theta = \begin{bmatrix} \Theta_{11} & \star & \star & \star & \star \\ \Theta_{21} & \Theta_{22} & \star & \star & \star \\ 6Q & 6Q & -12Q & \star & \star \\ \Theta_{41} & 0 & 0 & -\Omega & \star \\ \nabla^2\mathcal{H}_t A & \nabla^2\mathcal{H}_t A_d & 0 & \nabla^2\mathcal{H}_t A_e & \frac{Q^-1}{\delta_M}\end{bmatrix}
\end{equation}
and
\begin{align*}
    \Theta_{11} &= \frac{1}{2}{A+A^\top} + A^\top\nabla^2\mathcal{H}_t^\top P + P\nabla^2\mathcal{H}_t A - 4Q,\\
    \Theta_{21} &= A_d^\top\nabla^2\mathcal{H}_t^\top P + \frac{A_d^\top}{2} - 2Q,\\
    \Theta_{22} &= - 4Q + \sigma \mathcal{G}^\top \Omega \mathcal{G},\\
    \Theta_{41} &= A_e^\top\nabla^2\mathcal{H}_t^\top P + \frac{A_e^\top}{2}.
    \end{align*}
Besides, Assumption \ref{assu3} implies that $A$, $A_d$ and $A_e$ are constant matrices. Then, because of Assumption \ref{assu4}, constant matrices $\mathcal{H}_j$ with $j=1,...,N$ exist such that for all $\xi$ in $\mathbb{P}$ constants $0\leq \lambda_j \leq 1$ with $\displaystyle{\sum_{j=1}^{N}\lambda_j=1}$ such that
\begin{equation} \label{eq:cond_poly}
    \nabla^2\mathcal{H}_t = \sum_{j=1}^{N}\lambda_j \mathcal{H}_j.
\end{equation}
Note that $\nabla^2\mathcal{H}_t$ is affine with respect to the vertices of the polytope and, therefore, \eqref{eq:Theo_Schur} is also affine with respect to the vertices. Thus, $\Theta\prec0$ is satisfied if it is satisfied in the vertices of the polytope and the corollary is proved.
\end{proof}

\begin{remark}
    Note that the conditions in Theorem \ref{theo2} are state independent, but they are not a set of LMIs since \eqref{eq:Theo_Schur} depends on $Q^{-1}$. An easy approach to convert it into a LMI is to add a new constraint $Q\succ \alpha\mathcal{I}$, and consequently $Q^{-1}\prec\frac{\mathcal{I}}{\alpha}$.
\end{remark}

Besides, from Theorem \ref{theo1}, it is possible to easily derive conditions for interconnected linear port-Hamiltonian systems with delays. For linear port-Hamiltonian systems, Assumption \ref{assu3} is inherently satisfied, while the total energy of the interconnected system is $\mathcal{H}=\frac{1}{2}\xi^\top M \xi$ with $M\succ0$ to fulfill Assumption \ref{assu2}. Consequently, $\nabla_{t}\mathcal{H}=M\xi(t)$, and \eqref{eq:xi} is transformed into
\begin{equation} \label{eq:xi_linear}
    \dot{\xi}(t)=AM\xi(t)+A_dM\xi(t-\delta(t))+A_ee(t-\delta(t))
\end{equation}
Then, the following corollary is stated.

\begin{corollary}
For given positive scalars $\sigma$, $h$ and $\tau_M$, the interconnected delayed port-Hamiltonian system \eqref{eq:xi_linear} with event-triggered mechanism \eqref{eq:ETM} and under Assumptions \ref{assu1}-\ref{assu2} is globally asymptotically stable, if there exist matrices $P$, $Q$ and $\Omega$ - of appropriate dimensions - such that the set of LMIs
    \begin{equation} \label{eq:P_linear}
       (i)\,\, P \succ 0,\,\, (ii)\,\,  Q \succ 0,\,\ (iii)\,\,  \Omega \succ 0
    \end{equation}
    \begin{equation} \label{eq:Cor_linear}
        \Xi_l = \begin{bmatrix} {\Xi_l}_{11} & \star & \star & \star \\ {\Xi_l}_{21} & {\Xi_l}_{22} & \star & \star \\ 6Q & 6Q & -12Q & \star \\ {\Xi_l}_{41} & {\Xi_l}_{42} & 0 & {\Xi_l}_{44}\end{bmatrix}\prec0,
    \end{equation}
    is feasible with
    \begin{align*}
        {\Xi_l}_{11} &= \frac{1}{2}\left(A+A^\top\right) + A^\top M P + P M A\\ &+ \delta_M^2 A^\top M Q M A - 4Q,\\
        {\Xi_l}_{21} &= A_d^\top M P + \frac{A_d^\top}{2} \\& + \delta_M^2 A_d^\top M Q M A - 2Q,\\
        {\Xi_l}_{22} &= \delta_M^2 A_d^\top M Q M A_d - 4Q + \sigma \mathcal{G}^\top \Omega \mathcal{G},\\
        {\Xi_l}_{41} &= A_e^\top M P + \frac{A_e^\top}{2} + \delta_M^2 A_e^\top M Q M A,\\
        {\Xi_l}_{42} &= \delta_M^2 A_e^\top M Q \nabla^2\mathcal{H}_t A_d,\\
        {\Xi_l}_{44}&=\delta_M^2 A_e^\top M Q M A_e - \Omega .
        \end{align*}
\end{corollary}
\begin{proof}
    The proof is equivalent to the proof of Theorem \ref{theo1} but replacing $\nabla^2\mathcal{H}_t$ by $M$. Finally, since $A$, $A_d$, $A_e$ and $M$ are constant matrices by the definition of linear port-Hamiltonian system, it is shown that \eqref{eq:Cor_linear} is a LMI and the proof is completed. 
\end{proof}

\begin{remark}
    We consider the case with an ETM located after the output of $\Sigma_1$. However, it is possible to place it after $\Sigma_2$ or both of them, even considering different sampling periods \cite{Aranda2018}. The theoretical design is similar but considering that terms for the new error and the new delay would appear in \eqref{eq:xi}. So, the functional \eqref{eq:LKF} is also modified with a new term equivalent to $V_2$.
\end{remark}

%\textcolor{red}{Corollaries: CETC, nondelay or constant delay, linear port-hamiltonian systems. Global stability. Two ETMs. Casimirs?}

\section{Simulation results}

In this section, we provide simulations results to verify the results derived in the previous section. Let us consider a dumped normalized pendulum as described in \cite{garcia2005control,kao2012stability,van2014port}. The equations of the system are
\begin{equation} \label{eq:pend}
    \begin{aligned}
        &\ddot{q} + \sin(q) + \zeta\dot{q} = u
        & y_1 = \dot{q},
    \end{aligned}
\end{equation}
where $q$ is the angle described by the pendulum and $\zeta>0$ a damping constant. Considering the total energy of the system $H_1\left(q,\dot{q}\right)=\frac{1}{2}\dot{q}^2+\left(1-\cos(q)\right)$, \eqref{eq:pend} can be written in the port-Hamiltonian form \eqref{eq:pHsystem_i} with
\begin{equation}
        J_1=\begin{bmatrix}0 & 1 \\ -1 & 0\end{bmatrix}, \;
        R_1=\begin{bmatrix}0 & 0 \\ 0 & \zeta\end{bmatrix}, \;
        G_1=\begin{bmatrix}0 \\ 1\end{bmatrix}.
\end{equation}
Note that the origin is a stable equilibrium point, but not asymptotically stable. Therefore, we consider a controller with Hamiltonian $H_2\left(x_2\right)=\frac{K}{2}x_2^2$ such that $J_2=0$, $R_2=d_c$ and $G_2=1$, with $K>0$ a feedback gain $\zeta_c>$ a damping constant for the controller to be chosen. So, the whole system can be written in the form of \eqref{eq:xi} with $\mathcal{H}\left(\xi\right)=\frac{1}{2}\xi_2^2+\left(1-\cos(\xi_1)\right)+\frac{1}{2}\xi_3^2$ and
\begin{equation}
    A = \begin{bmatrix}0 & 1 & 0 \\ -1 & -\zeta & -1 \\ 0 & 0 & -\zeta_c\end{bmatrix}, \;
    A_d = \begin{bmatrix}0 & 0 & 0 \\ 0 & 0 & 0 \\ 0 & 1 & 0\end{bmatrix}, \;
    A_e = \begin{bmatrix}0 \\ 0 \\ 1\end{bmatrix}.
\end{equation}
Assumption \ref{assu1}-\ref{assu3} can be easily verified. So, to apply Theorem \ref{theo2}, we compute the Hessian of the total Hamiltonian $\nabla^2\mathcal{H}(\xi)=\text{diag}\left(\cos(\xi_1),1,K\right)$. Clearly, $-1\leq \cos(\xi_1) \leq 1$ and $\nabla^2\mathcal{H}(\xi)$ can be written in the form of \eqref{eq:cond_poly} with vertices $\mathcal{H}_1=\text{diag}\left(-1,1,K\right)$ and $\mathcal{H}_2=\text{diag}\left(1,1,K\right)$, so Assumption \ref{assu4} is verified. Then, using Theorem \ref{theo2}, the problem \eqref{eq:P}-\eqref{eq:Theo} is solved using LMI solvers. Let us fix $\zeta=0.1$, and $\zeta_c=1$ and $K=3$ for the controller. Note that these values can be adjusted or tuned with an emulation-based approach. Then, the solution of \eqref{eq:P}-\eqref{eq:Theo} depends on $\sigma$ and on $\delta_M=h+\tau_M$, i.e. the flexibility to trigger the events and the sum of the sampling period and the maximum admissible delay. Hence, a trade-off between both quantities is obtained and summarized in Table \ref{tab:pend_d_sigma}.

\begin{table}
\begin{center}
\label{tab:pend_d_sigma}\caption{Maximum values for $\delta_M$ and $\sigma$ which solve \eqref{eq:P}-\eqref{eq:Theo} for the normalized pendulum with $\zeta=0.1$, $\zeta_c=1$, $K=3$.}
\begin{tabular}{|c || c | c | c | c | c | c | c|} 
 \hline
 $\boldsymbol{\delta_M}$ & 0.1 & 0.2 & 0.3 & 0.4 & 0.5 & 0.6 & 0.7 \\ [0.5ex] 
 \hline\hline
 $\boldsymbol{\sigma}$ & 2.19 & 1.41 & 0.88 & 0.54 & 0.30 & 0.13 & 0.008 \\  [1ex] 
 \hline
\end{tabular}
\end{center}
\end{table}

\begin{figure}
    \centering
    \includegraphics[width=0.8\linewidth]{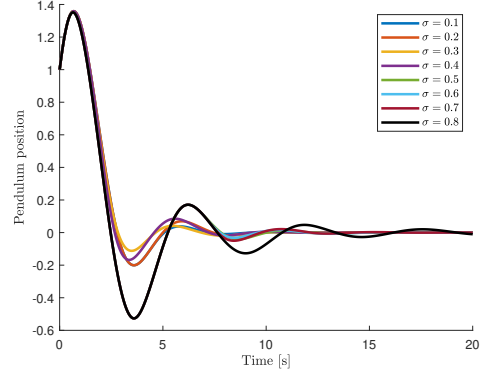}
    \caption{Pendulum position for $h=0.3$ and different values of $\sigma$.}
    \label{fig:pos_sigma}
\end{figure}

\begin{figure}
    \centering
    \includegraphics[width=0.8\linewidth]{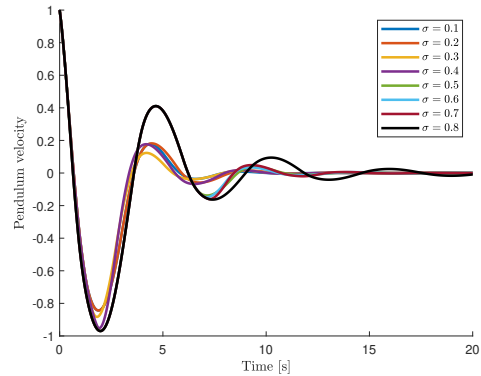}
    \caption{Pendulum velocity for $h=0.3$ and different values of $\sigma$.}
    \label{fig:vel_sigma}
\end{figure}

\begin{figure}
    \centering
    \includegraphics[width=0.8\linewidth]{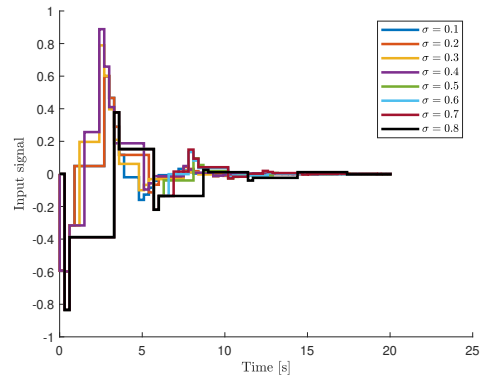}
    \caption{Input signal for $h=0.3$ and different values of $\sigma$.}
    \label{fig:input_sigma}
\end{figure}

To check the results, we have performed several simulations. First, we consider the case without delay, i.e. $h=\delta_M$, and set $h=0.3$. In Table \ref{tab:sigma}, the average inter-event time for different values of $\sigma$ and the Integral Square Error (ISE), Integral Absolute Error (IAE) and Integral Time Absolute Error (ITAE) of $H_1$ are summarized. It is shown that, in general, a larger $\sigma$ implies less triggered events, but some degradation in the performance since the input signal is not updated so often. However, these conclusions are not necessary. For example, for $\sigma=0.7$, more events are triggered than for $\sigma=0.6$. This might occur because not triggering an event in a certain instant might produce additional events in future samplings. In Figures \ref{fig:pos_sigma}-\ref{fig:input_sigma}, the position and velocity of the pendulum, and the input signal for the different values of $\sigma$ are depicted. As suggested by Table \ref{tab:sigma}, a lower $\sigma$ is related to a faster convergence to the equilibrium, but this is not a necessary consequence if avoiding updating the input signal implies that the pendulum is pushed in the correct direction with a larger torque for a larger time interval.

\begin{table*}
\begin{center}
\caption{Average inter-event times and indices ISE, IAE and ITSE for $h=0.3$ and different values of $\sigma$.}
\label{tab:sigma}
\begin{tabular}{|c || c | c | c | c | c | c | c| c|} 
 \hline
 $\boldsymbol{\sigma}$ & 0.1 & 0.2 & 0.3 & 0.4 & 0.5 & 0.6 & 0.7 & 0.8 \\ [0.5ex] 
 \hline\hline
 \textbf{Inter-event time} & 0.48 & 0.61 & 0.61 & 0.63 & 0.83 & 1.00 & 0.80 & 1.54 \\
 \hline
 \textbf{ISE} & 1.31 & 1.31 & 1.32 & 1.36 & 1.43 & 1.43 & 1.43 & 1.43 \\
 \hline
 \textbf{IAE} & 1.87 & 1.87 & 1.81 & 1.88 & 2.27 & 2.27 & 2.27 & 2.30 \\
 \hline
 \textbf{ITAE} & 2.27 & 2.31 & 2.08 & 2.27 & 3.81 & 3.80 & 3.83 & 4.08 \\[1ex] 
 \hline
\end{tabular}
\end{center}
\end{table*}

\begin{figure}
    \centering
    \includegraphics[width=0.8\linewidth]{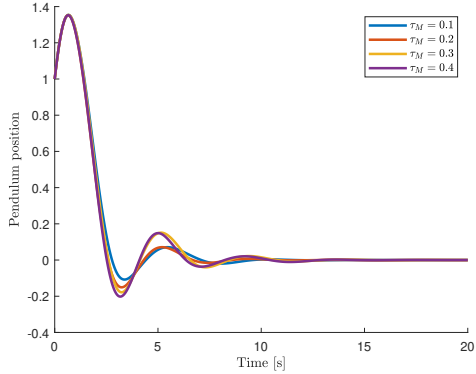}
    \caption{Pendulum position for $h=0.2$, $\sigma=0.2$ and different values of $\tau_M$.}
    \label{fig:pos_delay}
\end{figure}

\begin{figure}
    \centering
    \includegraphics[width=0.8\linewidth]{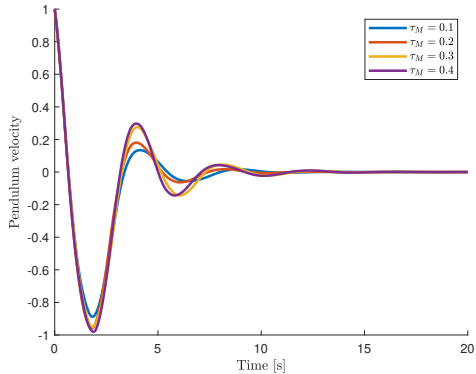}
    \caption{Pendulum velocity for $h=0.2$, $\sigma=0.2$ and different values of $\tau_M$.}
    \label{fig:vel_delay}
\end{figure}

\begin{figure}
    \centering
    \includegraphics[width=0.8\linewidth]{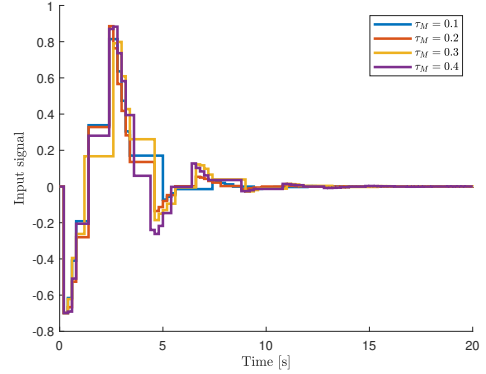}
    \caption{Input signal for $h=0.2$, $\sigma=0.2$ and different values of $\tau_M$.}
    \label{fig:input_delay}
\end{figure}

Finally, we test the system with parameters $h=0.2$, $\sigma=0.2$ and under a random time-varying delay $0<\tau(t)\leq \tau_M$. For the test, we simulate the system with different maximum delays. The average inter-event times and the measurements of indices ISE, IAE and ITSE are summarized in Table \ref{tab:delay}. There is a clear (and logical) correlation between the delay and the performance, which is more affected the larger the delay is. Besides, we can observe that a larger delay implies also more triggered events to compensate the effect. In Figures \ref{fig:pos_delay}-\ref{fig:input_delay}, we depict the temporal evolution of the system, which might be considerably affected by the delay. For example, the overshoot in pendulum position and velocity in Figure \ref{fig:pos_delay} and \ref{fig:vel_delay}, respectively, is clearly enlarged when the maximum delay is larger.

\begin{table}\label{tab:delay}
\begin{center}
\caption{Average inter-event times and indices ISE, IAE and ITSE for $h=0.2$, $\sigma=0,2$ and different values of $\tau_M$.}
\begin{tabular}{|c || c | c | c | c |} 
 \hline
 $\boldsymbol{\tau_M}$ & 0.1 & 0.2 & 0.3 & 0.4 \\ [0.5ex] 
 \hline\hline
 \textbf{Inter-event time} & 0.42 & 0.40 & 0.40 & 0.37 \\
 \hline
 \textbf{ISE} & 1.29 & 1.32 & 1.34 & 1.35 \\
 \hline
 \textbf{IAE} & 1.79 & 1.32 & 1.34 & 1.35 \\
 \hline
 \textbf{ITAE} & 2.05 & 2.09 & 2.39 & 2.40 \\[1ex] 
 \hline
\end{tabular}
\end{center}
\end{table}

\section{Conclusions}

In this paper, we have designed a periodic event-triggered strategy for the control of port-Hamiltonian systems. Besides, time-varying delays are included in the framework in order to increase the implementability in real environments. Stability conditions are derived using the Lyapunov-Krasovskii theory. Additionally, a mechanism is provided to convert these conditions into LMIs to facilitate numerical computations.

Several simulations have been carried out using a classical benchmark, a normalized damped pendulum. These simulations show, first, the benefits of the event-triggering strategy for a port-Hamiltonian system and, second, the existing trade-off among the parameters of the triggering condition, the sampling period, the time delays, and the performance of the system.

%Future research lines include the deepening into the different strategies of event-triggered control and its application to port-Hamiltonian systems, especially under different usual phenomena in communication networks such as packet losses, limited bandwidth, or cyber-attacks.
The application of event-triggered control to port-Hamiltonian systems opens the possibility of new research lines. On the one hand, the method can be extended to a larger class of systems, e.g., port-Hamiltonian systems with intrinsic time-delay, i.e. when the delays appear in the port-Hamiltonian system itself and not due to the communication network. Besides, once an event-triggered control framework is established, new triggering conditions, such as dynamic ones, can be tested to improve the transmission rate. On the other hand, common problems in networked control systems and cyber-physical systems such as packet losses or cyber-attacks are still unexplored for port-Hamiltonian systems. Their study together with event-triggered control might increase the implementability of the port-Hamiltonian framework in networked scenarios. 

\section*{Acknowledgments}

This work was partially supported by the Agencia Estatal de Investigación (AEI) under Project PID2020-112658RB-I00/AEI/10.13039/501100011033 and by the Universidad Nacional de Educaci\'on a Distancia (UNED) under Projects 2021V/-TAJOV/001, 2021V/PUNED/008 and 2022V/PUNED/007. The present work is part of the mobility stay ``Jos\'e Castillejo'' CAS21/00108 of Ernesto Aranda-Escol\'astico.

\bibliographystyle{IEEEtran}
\bibliography{bib}

\end{document}